\documentclass[letterpaper,twocolumn,10pt]{article}
\usepackage{usenix}
\usepackage{ulem}

\usepackage{tikz}
\usepackage{amsmath}
\usepackage[dvipsnames]{xcolor}
\usepackage{url}
\usepackage{tabularx}
\usepackage{subcaption}
\usepackage{makecell}

\usepackage{tikz}
\usetikzlibrary{arrows.meta,positioning,fit}

\usepackage{filecontents}

\begin{document}

\date{}


\title {\Large \bf MCP-in-SoS: Risk assessment framework for open-source MCP servers}


\author{
{\rm Pratyay Kumar}\\
New Mexico State University\\
pratyay@nmsu.edu
\and
{\rm Miguel Antonio Guirao Aguilera}\\
New Mexico State University\\
guirao@nmsu.edu
\and
{\rm Srikathyayani Srikanteswara}\\
GenAI and ML Research Visionary \\
srikathyayani.srikanteswara@gmail.com
\and
{\rm Satyajayant Misra}\\
New Mexico State University\\
misra@nmsu.edu
\and
{\rm Abu Saleh Md Tayeen}\\
University of Hartford\\
tayeen@hartford.edu
}

\maketitle

\begin{abstract}
Model Context Protocol (MCP) servers have rapidly emerged over the past year as a widely adopted way to enable Large Language Model (LLM) agents to access dynamic, real-world tools. As MCP servers proliferate and become easy to adopt via open-source releases, understanding their security risks becomes essential for dependable production agent deployments. Recent work has developed MCP threat taxonomies, proposed mitigations, and demonstrated practical attacks. However, to the best of our knowledge, no prior study has conducted a systematic, large-scale assessment of weaknesses in open-source MCP servers. Motivated by this gap, we apply static code analysis to identify Common Weakness Enumeration (CWE) weaknesses and map them to common attack patterns and threat categories using the MITRE Common Attack Pattern Enumerations and Classifications (CAPEC) to ground risk in real-world threats. We then introduce a risk-assessment framework for the MCP landscape that combines these threats using a multi-metric scoring of likelihood and impact. Our findings show that many open-source MCP servers contain exploitable weaknesses that can compromise confidentiality, integrity, and availability, underscoring the need for secure-by-design MCP server development.
\end{abstract}
\section{Introduction}\label{sec:Introduction}

Model Context Protocol (MCP)~\cite{anthropic2024mcp} is an open standard, introduced by Anthropic in November 2024, for connecting LLM-based assistants to external tools and data sources through a uniform, model-agnostic interface.\footnote{https://www.anthropic.com/news/model-context-protocol} In little over a year, MCP has seen rapid ecosystem adoption: public registries and community-curated collections track thousands of MCP server repositories, and industry reporting suggests MCP deployments now span well into the five-figure range.\footnote{https://mcp.so/}

This growth is driven by MCP's clean separation of concerns. The LLM client performs task reasoning and selects capabilities, while MCP servers expose structured tools/resources with explicit schemas and standardized invocation semantics. Compared to earlier ad-hoc agent integrations (hard-coded wrappers, prompt-embedded tool instructions, vendor-specific plugins, or fragile function-calling conventions), MCP provides a stable contract that improves portability across models and simplifies operationalization.

At the same time, MCP changes the security posture of agentic systems. Traditional client-server applications typically have fixed, pre-programmed call graphs and relatively static workflows. In MCP-powered systems, control flow is orchestrated at runtime by an agent’s reasoning loop: the client can chain tool calls, synthesize arguments dynamically, and trigger server-side actions that transitively reach sensitive resources (filesystem operations, network access, credentialed APIs, developer tooling, and downstream services). As a result, commonplace implementation flaws can amplify beyond a single request boundary, enabling cross-tool escalation paths and increasing the radius of compromise.

Recent work has proposed MCP threat taxonomies~\cite{hou2025model}, demonstrated practical attacks~\cite{guo2025systematic}, and discussed mitigations~\cite{kumar2025mcp}. However, the ecosystem is currently dominated by open-source MCP servers that are routinely cloned and deployed, and there is limited evidence on how weaknesses manifest in the code at scale across real implementations. This gap matters because MCP servers sit at the boundary between an agent and high-privilege capabilities, making them a natural choke point for both prevention and governance. 
To systematically characterize these implementation-level risks, there is a need for a standardized pipeline that captures recurring software weaknesses and their exploitation pathways. Common Weakness Enumeration (CWE) provides a structured vocabulary for identifying and comparing recurring code-level weaknesses, while CAPEC models attacker-facing exploitation patterns that arise from these weaknesses. Leveraging both CWE and CAPEC can enable comparable analysis that links implementation flaws in MCP servers to practical attack behaviors.

To address this gap, we introduce
\textbf{MCP-in-SoS}, a structured pipeline for detecting and prioritizing security weaknesses in MCP server implementations. Our four-stage pipeline operates by (i) extracting candidate security weaknesses from MCP server implementations; (ii) normalizing and mapping these findings to standardized CWE classes; (iii) linking the identified CWE classes to CAPEC attack patterns to compute metadata-driven risk scores; and (iv) correlating the resulting vulnerabilities with MCP-specific threat surfaces (Protocol, Tool, Resource, and Prompt) to quantify their conditional co-occurrence, thereby revealing empirically supported multi-stage exploit chains.

Using \textbf{MCP-in-SoS}, we conduct a large-scale analysis of 222 GitHub repositories implementing MCP servers. Our measurements show that security weaknesses are widespread and frequently compositional in nature. Under our methodology, 191 of the 222 repositories (86.0\%) contain at least one mapped weakness, highlighting the substantial security exposure in the current MCP ecosystem. Our proposed CWE-CAPEC–driven scoring mechanism indicates that the majority of the analyzed repositories fall into the elevated-risk bands (High or Very High). Finally, our conditional co-occurrence analysis shows that MCP threat surfaces rarely manifest in isolation; instead, Protocol weaknesses frequently coincide with Tool, Resource, and Prompt vulnerabilities, resulting in recurring, multi-stage exploitation chains in real-world deployments.

We make the following contributions:
\begin{itemize}
  \item A large-scale weakness assessment of 222 publicly available Python MCP server repositories from GitHub.
  \item A reproducible pipeline that combines static analyzers with an LLM-assisted check to extract, normalize, and map findings to CWE classes.
  \item A set of Joern~\cite{joern} queries aligned with the 2025 CWE Top 25 for detecting high-impact weakness patterns in Python MCP servers.
  \item A risk scoring model driven by CWE-CAPEC metadata, producing both CWE-level and repository-level risk metrics.
  \item An MCP-aligned threat-surface taxonomy and a conditional co-occurrence analysis that reveals common multi-stage 
  exploit chains.
\end{itemize}

Sections~\ref{sec:related_work} and~\ref{sec:background} cover background and related work. Section~\ref{sec:threat_model} defines our threat model, Section~\ref{sec:methodology} describes our methodology, and Section~\ref{sec:evaluation} presents results. Section~\ref{sec:conclusion} concludes with findings and limitations.
\section{Related Work}
\label{sec:related_work}

\subsection{Large Language Model Agents}
\label{sec:llm-agent}
Large language models (LLMs) are Transformer-based neural networks trained on massive text corpora to predict the next token in a sequence. Modern LLMs, such as GPT-4~\cite{achiam2023gpt}, Claude~\cite{anthropic_claude_2025}, and Gemini~\cite{google_gemini_2025}, contain tens of billions to trillions of parameters and generate outputs token by token based on learned probability distributions, with capabilities emerging from training rather than explicit programming.
Building on LLMs, the shift from conversational models to autonomous AI agents marks a key architectural change. Unlike stateless, reactive chat models, AI agents perceive environments, reason, decide, and act toward explicit goals~\cite{DBLP:journals/corr/abs-2504-15585}. Enabling LLMs to invoke external functions, such as web search, code execution, or IoT control, extends them from passive text generators to general-purpose platforms that integrate reasoning with real-world actions~\cite{li2025review}.

Recent work on LLM-based agents spans both single-agent and multi-agent systems~\cite{cheng2024exploring}, with a strong focus on enhancing reasoning, planning, tool use, and long-term memory~\cite{du2025survey}. LLM agents have been applied across domains such as software engineering~\cite{He2024LLM-Based}, scientific discovery~\cite{Ramos2024A}, simulation~\cite{Gao2023Large}, and industry~\cite{Guo2024Large}, alongside the development of benchmarks and evaluation frameworks for web tasks, tool use, multi-step reasoning, and multi-agent coordination~\cite{Ferrag2025From}.


\subsection{MCP Ecosystem Security}
\label{sec:mcp-security}
As MCP continues to attract increasing attention and adoption, several studies have begun to examine its associated security challenges from different perspectives. 
Hou et al.~\cite{hou2025model} provided a comprehensive threat taxonomy covering diverse attacker types and validated it through empirical case studies. Their study assessed the MCP ecosystem, identified key limitations, and offered practical recommendations to improve security and governance, while outlining future directions to strengthen MCP resilience.
Kumar et al.~\cite{kumar2025mcp} proposed MCP Guardian to mitigate security risks in MCP deployments by incorporating authentication, rate limiting, Web Application Firewall (WAF) scanning, and comprehensive logging, all while preserving the simplicity of the MCP workflow. 

Narajala and Habler~\cite{Narajala2025Enterprise-Grade} presented practical implementation strategies, operational guidelines, and security patterns for MCP adopters. Their framework supports systematic risk prioritization via structured threat modeling and qualitative assessment of attack vectors, providing a foundation for secure, enterprise-grade MCP deployment.
Fang et al.~\cite{fang2025we} introduced SAFE-MCP, a controlled framework for systematically studying safety issues in MCP-based agents. Their pilot experiments showed that these risks are concrete and difficult to mitigate, and they outlined key research directions, such as red teaming and MCP-aware safe LLMs to build a secure and resilient MCP ecosystem.
Guo et al.~\cite{guo2025systematic} introduced MCPLIB, a plugin-based attack simulation framework that categorizes 31 MCP attack types into four classes, including direct/indirect tool injection, malicious user attacks, and LLM-inherent weaknesses. MCPLIB enables the first quantitative evaluation of attack effectiveness and root-cause analysis of MCP vulnerabilities, revealing systemic weaknesses in MCP agent design and operation.

\subsection{MCP Server Security}
\label{sec:mcp-server-security}
Several studies have examined the security landscape of MCP, with a primary focus on vulnerabilities and threats associated with MCP servers.
Song et al.~\cite{Song2025Beyond} conducted an end-to-end empirical study of four attack classes in the MCP ecosystem. Their results show that existing auditing and user defenses are inadequate, and user studies reveal widespread difficulty in identifying malicious MCP servers, with attacks capable of causing serious local harm.
Zhao et al.~\cite{zhao2025mcp} analyzed MCP servers as active threat actors, decomposing their components to study how adversarial developers embed malicious behavior. They proposed a component-based taxonomy of twelve attack categories validated with proof-of-concept servers in real-world host–LLM settings, showing that malicious MCP servers are low-cost to create, hard to detect, and capable of causing tangible harm.

Wang et al.~\cite{wang2025mcpguard} systematically analyzed associated security threats, focusing on the architectural layer of MCP servers, and identified challenges such as code-level and traditional web vulnerabilities in
MCP servers.
Radosevich and Halloran~\cite{Radosevich2025MCP} introduced McpSafetyScanner, the first automated tool for assessing the security of arbitrary MCP servers. Using agent-based analysis of server tools, prompts, and resources, it detects vulnerabilities and provides remediation guidance. Their empirical results show accurate detection of known vulnerabilities and effective remediation recommendations.

While the prior works on MCP server security discussed above provide valuable insights, they predominantly focus either on theoretical analyses of MCP server vulnerabilities based on architectural patterns~\cite{wang2025mcpguard} or on constructing malicious MCP servers to launch and evaluate attacks against various MCP hosts and LLM models~\cite{Song2025Beyond,zhao2025mcp}. To the best of our knowledge, none of the existing studies have systematically assessed the security risks of publicly available open-source MCP servers, particularly with respect to critical software-level (code-level) vulnerabilities. In this paper, we address this gap by applying static analysis techniques, including CodeQL, to examine the prevalence and patterns of Common Weakness Enumeration (CWE) vulnerabilities in a collection of MCP server implementations mined from open-source repositories.




\section{Background}\label{sec:background}
\subsection{Model Context Protocol}
\label{sec:mcp}
The Model Context Protocol (MCP) is a general-purpose framework introduced by Anthropic~\cite{anthropic2024mcp} in late 2024. MCP is designed to standardize and formalize interactions between AI agents and external tools. It provides a unified framework enabling AI applications to dynamically discover, select, and coordinate external tools based on the user’s task context.
\begin{figure}[htbp]
\centerline{\includegraphics[width=0.47\textwidth]{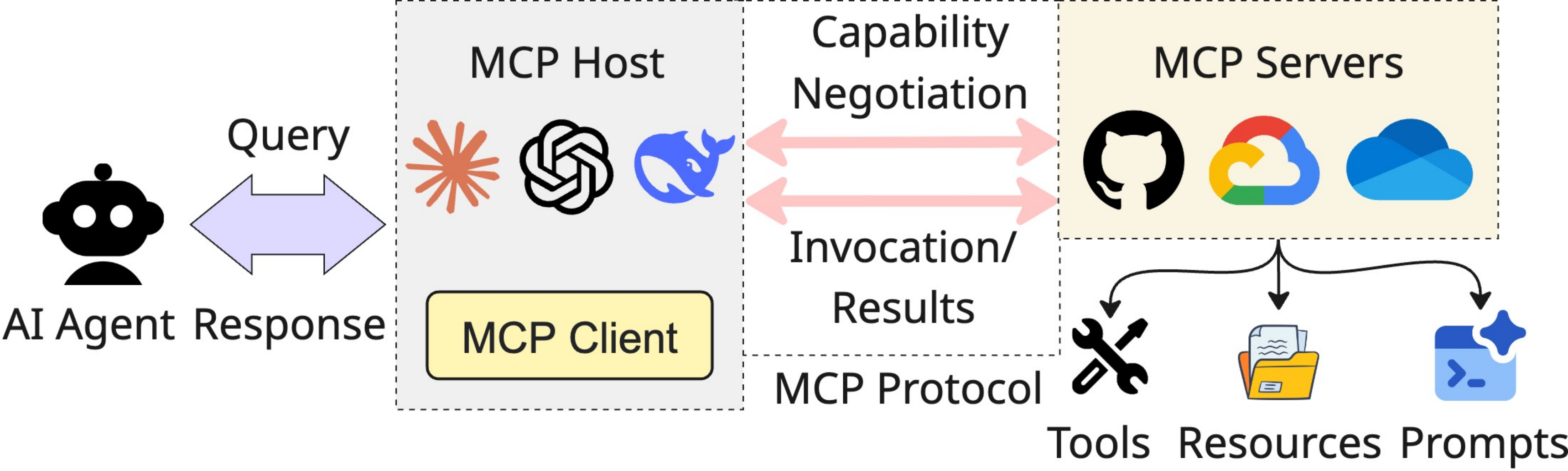}}
\caption{MCP architecture.}
\label{fig:mcp-arch}
\end{figure}

As illustrated in Figure~\ref{fig:mcp-arch}, 
MCP comprises three core components: the User/AI agent, the MCP Host (which encapsulates the LLM agent), and one or more MCP Servers. Initially, the MCP Host establishes connections with available MCP Servers to identify and assess their capabilities. These servers expose essential metadata, including tool names and natural-language descriptions, which are incorporated into the LLM agent’s context or system prompt during the initialization and registration phase. Upon receiving a user query, the agent determines the appropriate tool and issues a corresponding tool call to the relevant server. The server then executes the requested tool and returns the execution result to the agent. Finally, the agent integrates the returned output and generates a coherent final response for the user.

\subsection{Static Analysis Tools}
\label{sec:static-analysis-tools}
Static Analysis has been an integral part of Software Quality Assurance, as it provides extensive information about code flow and control structures, which are important for identifying vulnerabilities and increasing the robustness of software applications~\cite{chen2019enfuzz,hu2025sok,shi2025x,yamaguchi2014modeling}. Static Analysis offers many different techniques that are commonly used for vulnerability detection of codebases. These include Fuzzing\cite{hu2025sok}, Ensemble Fuzzing~\cite{chen2019enfuzz}, Concolic Execution~\cite{liu2024co3}, and Code Property Graphs (CPG)~\cite{yamaguchi2014modeling}.

Fuzzing is a popular software testing technique for vulnerability detection. The key idea of fuzzing is to generate plenty of inputs to execute the target application and monitor the results for any anomalies that may arise. Each fuzzing engine develops its own strategies to generate and apply the inputs into an application code~\cite{chen2025elfuzz}. Given the different strategies that each fuzzing engine may apply, a logical improvement is to combine fuzzers with different strategies into an Ensemble Fuzzing, in order to increase the coverage of input generation techniques~\cite{chen2019enfuzz}. Concolic Execution is also a prominent software testing technique that has proven to have success over traditional symbolic execution in software applications, especially for the detection of critical security vulnerabilities~\cite{liu2024co3}. Finally, CPG merges Abstract Syntax Trees (AST), Control Flow Graphs (CFG), and Program Dependency graphs (PDG)~\cite{yamaguchi2014modeling}. This comprehensive representation enables us to model the codebase as a path traversal along the CPG for detecting different vulnerabilities. 


There are several open source tools that are available for static analysis. We mention three well-known ones, which are freely available. 
CodeQL~\cite{codeql} is a tool developed by GitHub, Inc., and it is integrated into the GitHub offering for public and open-source projects. The tool can be executed outside the GitHub ecosystem in a local environment. The goal of this tool is to allow the security analyst to query the codebase using a query language similar to SQL, treating it as data in a database, and search for vulnerable patterns. This query language provides great flexibility for searching for known vulnerable patterns as well as new patterns that the security analyst may want to find. 

Joern~\cite{joern} is a platform for robust source-code analysis. It transforms the codebase into a Code Property Graph (CPG), a graph representation of code that enables easy analysis. Once the code is represented as CPG, it can be queried for vulnerable patterns, general parsing, and taint analysis.
Cisco AI Defense MCP Scanner~\cite{mcp-scanner} is a Python tool for scanning and analyzing MCP servers and tools for potential security findings. This tool combines different other tools, such as Cisco AI Defense inspect API, YARA\footnote{YARA is a tool aimed at helping malware researchers to identify and classify malware samples.} rules, and its LLM-as-a-judge to detect malicious MCP tools. 

All of the above tools were integrated into our analysis pipeline to identify candidate vulnerabilities. 

\subsection{Reference Databases}
\label{sec:ref_db}

In this section, we define the two main databases that will serve as references during our experiments: the Common Weakness Enumeration (CWE)\footnote{https://cwe.mitre.org/data/index.html} and the Common Attack Pattern Enumeration and Classification (CAPEC)\footnote{https://capec.mitre.org/index.html} databases. In Table~\ref{tab:placeholder_label}, the elements correspond to weaknesses for CWE, and for CAPEC, the elements correspond to attacks.  

\begin{table}[h]
    \centering
    \caption{CWE and CAPEC database details. }
    \label{tab:placeholder_label}
\begin{tabular}{|l|c|l|}
\hline
\textbf{Database Name} & \textbf{Number of Elements} & \textbf{Version} \\ \hline
CWE & 399 & 4.19.1 \\ \hline
CAPEC & 615 & 3.9 \\ \hline
    \end{tabular}
\end{table}

\subsubsection{CWE Database}
The CWE database contains a list of software and hardware weaknesses. Given our focus on the MCP Server only, we focus on the subset of software development weaknesses. After careful examination and analysis of the database's XML schema, we identified several risk factors to be integrated into a Risk Index (RI) formula. 
We provide the following definitions for each risk factor in the CWE database. 

\textbf{Weakness:} A weakness ($w$) is a mistake or condition that, if left unaddressed, could, under the proper conditions, render a cyber-enabled capability vulnerable to attack, allowing an adversary to cause items to function in unintended ways.

\textbf{Likelihood of Exploit (LE)}. This value assesses the likelihood that the weakness can be successfully exploited by an attacker. 

\textbf{Common Consequences (CC)}. These are used to specify individual consequences associated with a weakness. These individual consequences are Impact types that could materialize when the weakness is exploited.

\textbf{Mode of Introduction (MI)}. The mode of introduction provides information on how and when a given weakness may be introduced into the software. 

\subsubsection{CAPEC Database}
The CAPEC database provides a publicly available catalog of common attack patterns that helps users understand how adversaries exploit weaknesses in applications and other cyber-enabled capabilities. We provide the following definitions for each of the risk factors from the CAPEC database.
 
\textbf{Attack Pattern (AP)}. These are descriptions of the common attributes and approaches adversaries use to exploit known weaknesses in cyber-enabled capabilities. 

\textbf{Likelihood of Attack (LA)}. This risk factor captures the average likelihood that an attack leveraging this attack pattern will succeed. 

\textbf{Typical Severity (TS)}. This risk factor captures an overall average severity value for attacks that leverage this attack pattern.


\section{System and Threat Model}
\label{sec:threat_model}

\begin{figure*}[htbp]
\centerline{\includegraphics[width=1.0\textwidth]{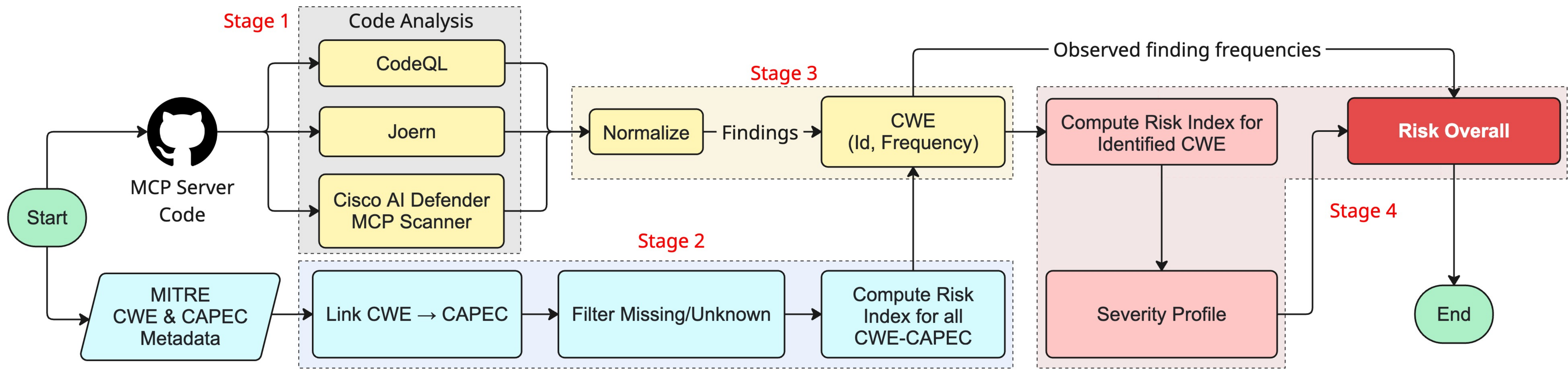}}
\caption{Overview of our four-stage framework for analyzing MCP server repositories and computing risk scores.}
\label{fig:pipeline}
\end{figure*}

\subsection{System Model}
We consider an MCP deployment consisting of an MCP server and one or more MCP clients that invoke server-exposed tools over the MCP transport. In many real deployments, the client is an LLM-based agent (or an application embedding an LLM), but our analysis does not depend on any particular client design; we treat the client as an untrusted requester and focus on weaknesses in the MCP server implementation.

The MCP server runs with the privileges of its host process and may have access to sensitive resources, including local files, environment variables, and secrets, network connectivity, and credentialed downstream services (e.g., internal APIs or cloud services). We treat the MCP server as the security boundary: any compromise of confidentiality, integrity, or availability within the server context (or the resources it can reach) is in scope. If the server implements authentication or authorization, these mechanisms are part of the server and are in scope for correctness.


\subsection{Threat Model}
The adversary's goal is to misuse an agent to exploit server-side tools and vulnerabilities to exfiltrate sensitive information. Specifically, the attacker can use our analysis pipeline to detect vulnerabilities in MCP server implementations, then craft an adversarial context to direct the agent to invoke vulnerable tools in ways that trigger exploitation. Model-side and user-side behaviors are out of scope, including prompt-injection effects on an LLM client's reasoning and attacks requiring prior compromise of the underlying cloud provider, independent of an MCP server weakness.



    

\section{Methodology}
\label{sec:methodology}
\subsection{Framework Overview}
\label{sec:framework-overview}
While the MCP protocol specification provides architectural foundations for secure AI-to-tool interactions, actual server implementations can systematically violate fundamental security principles. Given the rapid proliferation of thousands of diverse, independently developed (open-source) MCP servers, manual security auditing at the ecosystem scale is infeasible, underscoring the need for a systematic approach to assess MCP server security at the implementation level. Motivated by this gap, we propose \textbf{MCP-in-SoS}, an automated analysis pipeline that systematically identifies and prioritizes security weaknesses in MCP server implementations.


Our pipeline, shown in Figure~\ref{fig:pipeline}, operates in four stages: (i) \textbf{Static Code Analysis}, 
which scans each MCP server repository with multiple analyzers to extract candidate weakness findings. (ii) \textbf{Metadata Preparation},
which ingests the MITRE CWE and CAPEC databases, filters missing/unknown fields, and precomputes Risk Index values for all usable CWE-CAPEC pairs. (iii) \textbf{Normalization}, 
which deduplicates and normalizes analyzer outputs and aggregates them into per-repository CWE identifiers with observed finding frequencies. And, (iv) \textbf{Risk Scoring}, which assigns each observed CWE its Risk Index, derives a repository severity profile, and combines severity with observed frequencies to compute an overall repository risk score. 

\subsection{Static Code Analysis}
\label{sec:static-code-analysis}

Static analysis is a core component of software assurance because it exposes control/data-flow structure without requiring execution, enabling systematic screening for vulnerability patterns at scale. In Stage~1 of our pipeline, shown in Figure~\ref{fig:pipeline}, we run three analyzers over each MCP server repository snapshot to extract candidate weakness findings: CodeQL~\cite{codeql}, Joern~\cite{joern}, and (on a subset) the Cisco AI Defender MCP Scanner~\cite{mcp-scanner}.

CodeQL constructs a code database for each repository and evaluates security queries against that database to emit alerts with source locations and (when supported by the query) data-flow traces. We use CodeQL to capture well-studied weakness patterns in Python projects, leveraging a declarative query model widely adopted in both research and practice. CodeQL’s strength in our setting is high-precision identification of known patterns (e.g., injection sinks/sources, unsafe deserialization, and access-control anti-patterns) with traceability to concrete program locations.

Joern parses each repository into a Code Property Graph (CPG), enabling expressive structural and data-flow queries over program elements. We use Joern to complement CodeQL by implementing MCP-relevant queries that naturally express graph structure (e.g., flows from request parameters to sensitive sinks). For our Python MCP server dataset, we develop a Python-oriented Joern query suite aligned with the 2025 CWE Top 25.\footnote{https://cwe.mitre.org/top25/archive/2025/2025\_cwe\_top25.html} We drafted the queries using Claude Sonnet 4.5~\cite{anthropic_claude_sonnet_4_5} and iteratively refined them by executing against a local Joern instance to ensure syntactic validity and stable behavior across repositories. The CWE-ID, Name, and Description are provided to the LLM to create an initial query for a specific CWE, then it is tested for syntax correctness against a Joern instance and cycle again in case of errors until a correct and error-free query results that can detect a specific CWE. This process creates a total of 54 queries that we use to detect multiple vulnerabilities during the Code Analysis stage in Figure~\ref{fig:pipeline}.

The output of this stage is a set of Joern matches with concrete code locations and query provenance, which we normalize and map to CWE identifiers in Stage~3 (Section~\ref{sec:normalization}).

We additionally use the Cisco AI Defender MCP Scanner as a secondary analysis component on a subset of repositories to complement CodeQL and Joern. The scanner integrates multiple detectors (including rule-based signatures and LLM-assisted triage) to flag MCP-relevant risky behaviors and malicious patterns that may evade purely static rule suites due to rule scope or tooling differences.

Across tools, Stage 1 produces a set of candidates. These candidates are normalized, mapped to CWE identifiers, and aggregated per-repository frequencies in Stage~3 before applying metadata-driven risk scoring in Stage~4.

\subsection{Metadata preparation}
\label{sec:cwe_capec_ins}


Stage~2 of our pipeline prepares the metadata used for risk scoring in Stage~4. In this context, metadata refers to the structured risk attributes that MITRE publishes for each CWE weakness and CAPEC attack pattern on the respective websites. We are particularly interested in the Likelihood of Exploit, Common Consequences, and Mode of Introduction from the CWE schema, and the Likelihood of Attack and Typical Severity from the CAPEC schema. These fields provide the inputs to our Risk Index formula in Section~\ref{sec:normalization}.

Preparation involves three steps. First, we ingest the full CWE and CAPEC databases and link each CWE entry to its related CAPEC attack patterns. Second, we perform an exploratory data analysis (EDA) on both databases to assess data coverage, identify missing values across risk factors, and observe trends in how weaknesses and attack patterns are distributed. Third, we filter out entries with missing or unknown required fields - keeping only CWE–CAPEC pairs where all scoring inputs are available, and compute Risk Index values for all usable pairs. The EDA findings that guide our filtering and imputation choices are reported in the following subsections.

\subsubsection{Common Weakness Enumeration}
\begin{figure}[!ht]
    \centering
    \includegraphics[width=0.95\columnwidth]{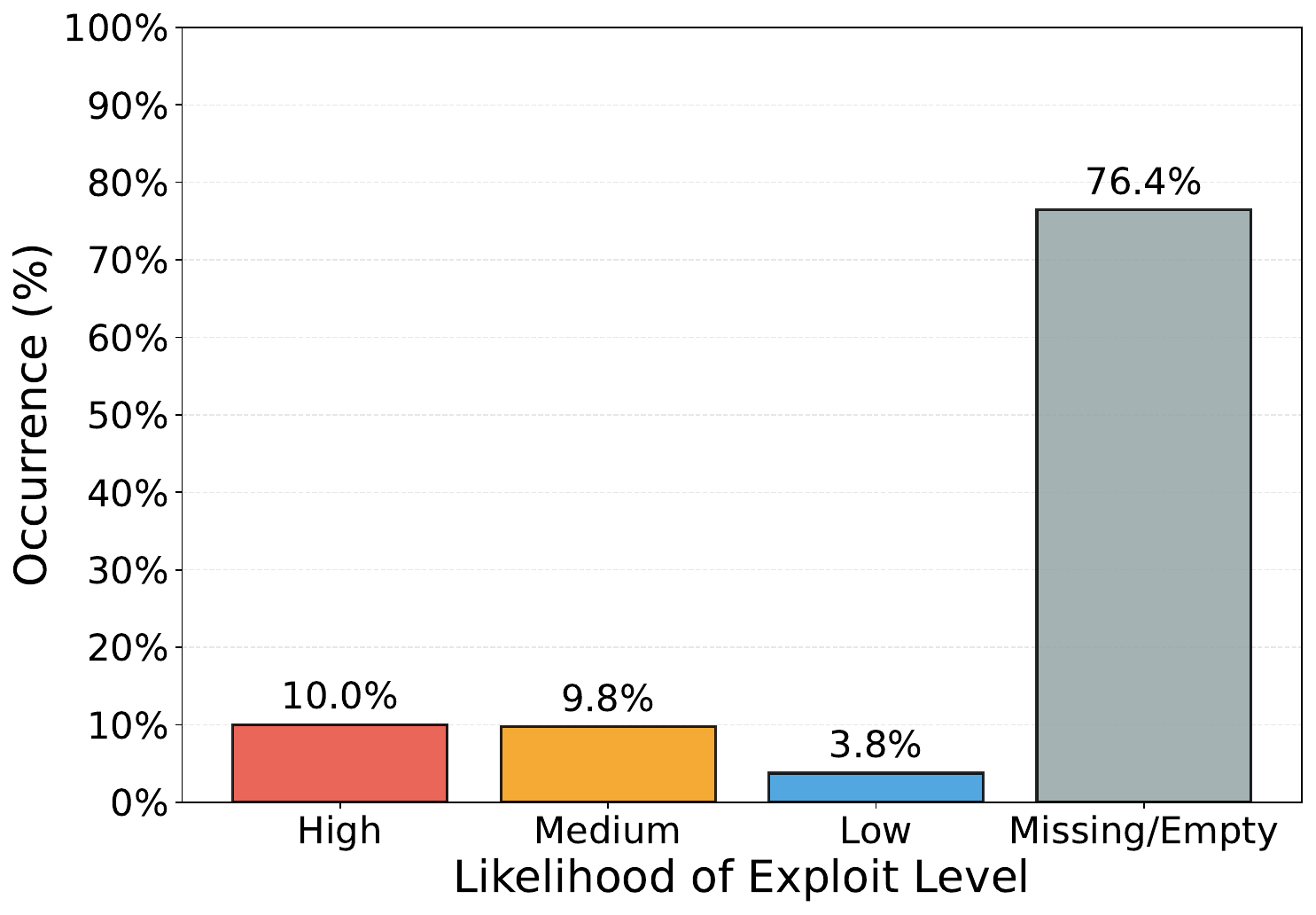}
    \caption{Likelihood of Exploit distribution. }
    \label{fig:le}
\end{figure}

\textbf{Likelihood of Exploit (LE):} The current state of the database \footnote{Both databases accessed on January 31, 2026.} shows that High and Medium likelihood of exploits account for nearly 10\% of the total weaknesses each. We find that 76.4\% (305) of the weaknesses have a missing or empty value. Figure \ref{fig:le} shows the distribution among the three categories and missing values. For each instance of a missing LE value, we select the Likelihood of Attack value from the CAPEC database by mapping the weakness to its related attack pattern. The inverse process is follow for the missing Likelihood of Attack values in the CAPEC database.

\textbf{Common Consequences (CC):} This risk factor details the different impacts (e.g., bypass protection) that the exploitation of a specific weakness may have in an application. The database contains a total of 867 impact elements, of which 25 are unique elements.
Table \ref{tab:top10_impacts} shows the top 5 impacts that occur more frequently among the weaknesses, from which we can observe that the first four impact elements are related to authentication and authorization issues.

\begin{figure}[!ht]
    \centering
    \includegraphics[width=0.95\columnwidth]{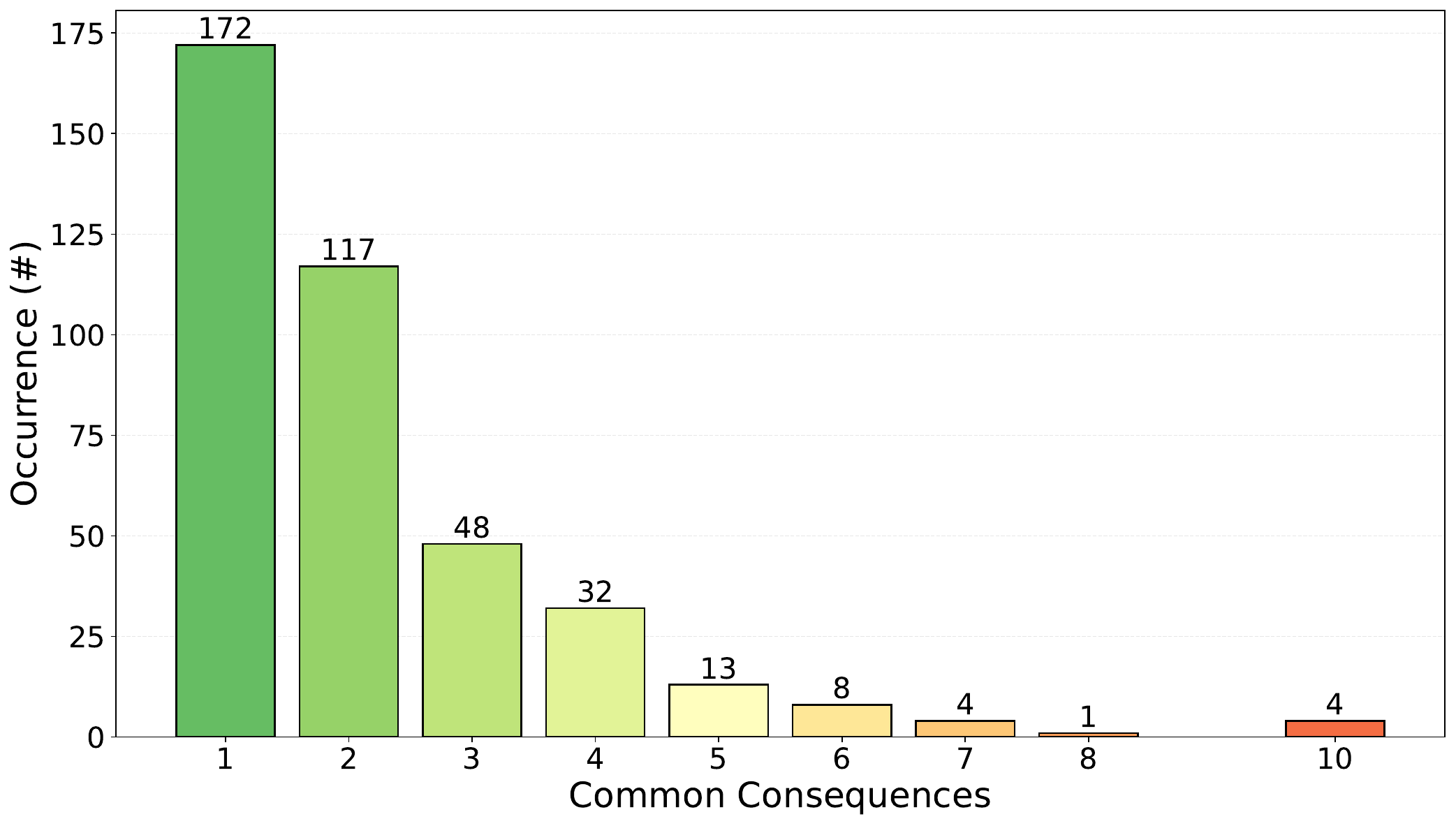}
    \caption{Number of Impact elements per weakness. }
    \label{fig:impact_per_weak}
\end{figure}

\begin{table}[]
    \centering
    \caption{Top 5 Impact elements in the CWE database.}
    \label{tab:top10_impacts}
\begin{tabularx}{\columnwidth}{|c|X|c|}
\hline
\textbf{Pos.} & \textbf{Impact} & \textbf{Occurrences} \\ \hline
1 & Bypass Protection Mechanism & 77 \\ \hline
2 & Read Application Data & 73 \\ \hline
3 & Execute Unauthorized Code or Commands & 68 \\ \hline
4 & Gain Privileges or Assume Identity & 67 \\ \hline
5 & DoS: Crash, Exit, or Restart & 58 \\ \hline
\end{tabularx}
\end{table}

We also look at this risk factor from the perspective of the number of Impact elements per weakness. Figure \ref{fig:impact_per_weak} shows that 43.1\% of the weaknesses have only one impact element, 72\% of the weaknesses have 1 and 2 impact elements, while the maximum number of impact elements per weakness is $10$, which happened only 4 times in the CWE database. This finding means that the number of impact elements is low among all weaknesses in the database.
Only 7.6\% of the weaknesses have more than 4 impacts, meaning that these cases have a more serious impact if the related weakness is exploited. We can interpret these values as the extent of damage that an exploited weakness can bring to the system. The mean is 2.17, the median is 2, and the standard deviation is 1.56 impact elements per weakness. 

\begin{figure}[!ht]
    \centering
    \includegraphics[width=0.95\columnwidth]{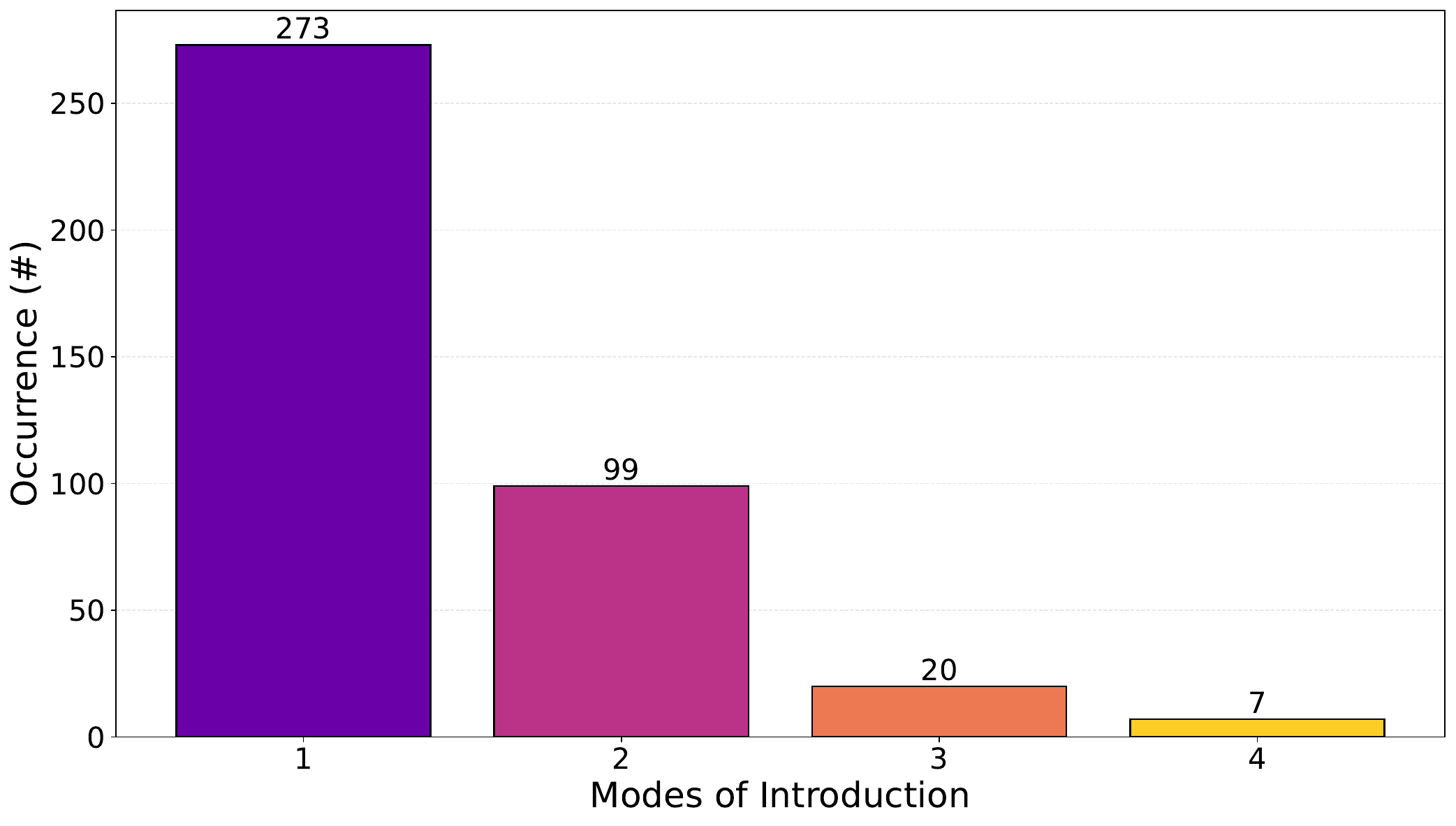}
    \caption{Modes of Introduction distribution. }
    \label{fig:mi}
\end{figure}

\textbf{Mode of Introduction (MI)}. A low number of MI indicates that there are few areas where a weakness can be introduced into the software, and hence the likelihood of compromise is low. On the contrary, a high number means that there are many more areas where that weakness can be introduced. In this analysis, we find that 68.4\% of the weaknesses have only one MI. Only 6.8\% of the weaknesses have either 3 or 4 MIs, indicating that attackers will have a small attack surface when exploiting these weaknesses. The distribution is left-skewed, with very few weaknesses having 3 or 4 modes. The MI has a mean of 1.4, a median of 1.0, and a standard deviation of 0.67 phase elements per weakness. The low standard deviation (0.67) indicates that most weaknesses cluster around one phase.  Refer to Figure \ref{fig:mi} for a detailed description.

\subsubsection{CAPEC Database}
The CAPEC database also has several features. 

\begin{figure}[!ht]
    \centering
    \includegraphics[width=0.95\columnwidth]{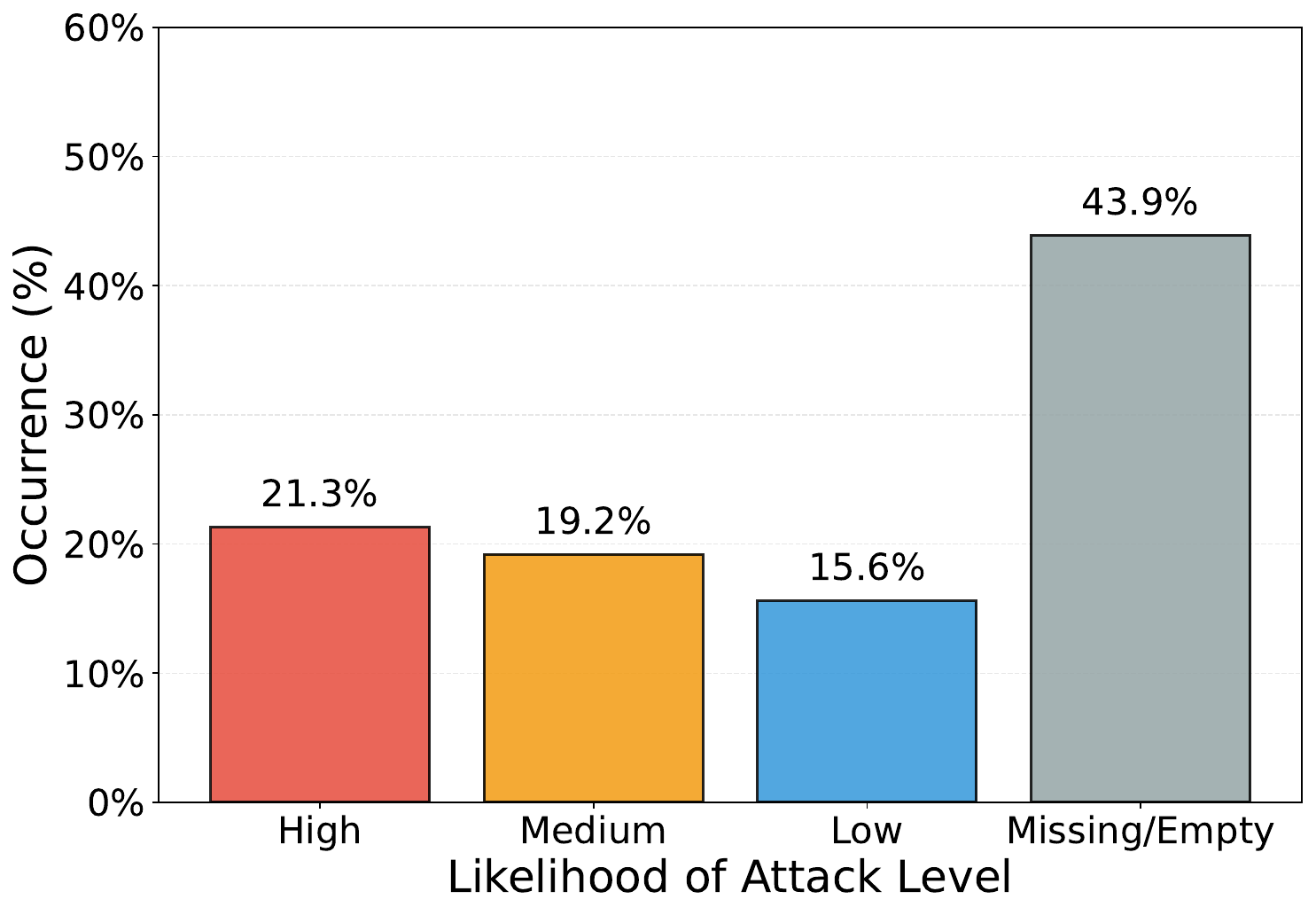}
    \caption{Likelihood of Attack distribution. }
    \label{fig:la}
\end{figure}

\textbf{Likelihood of Attack (LA).} This feature refers to the risk factor. It shows that 43.9\% of the attack patterns have missing or empty likelihood-of-attack data. For each instance of a missing LA value, we select the Likelihood of Exploit value from the CWE database by mapping the weakness to its related attack pattern. The cases where both Likelihood values are missing are discarded, as we lack information to evaluate the risk. Figure~\ref{fig:la} shows the missing proportion while showing the proportion in the High, Medium, and Low categories.

\begin{figure}[!ht]
    \centering
    \includegraphics[width=0.95\columnwidth]{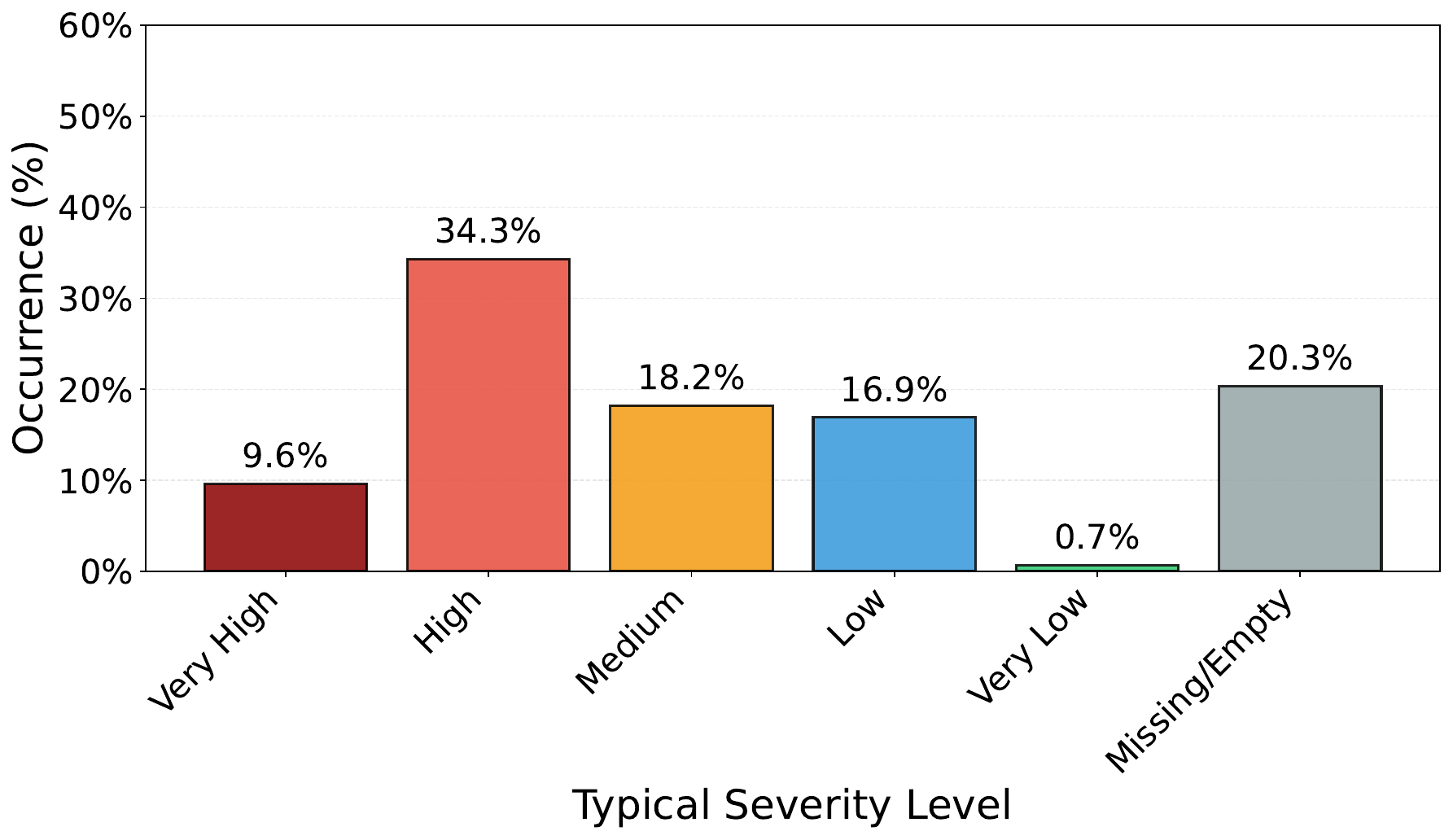}
    \caption{Typical Severity distribution. }
    \label{fig:ts}
\end{figure}

Comparing CWE's Likelihood of Exploit versus CAPEC's Likelihood of Attack, we found that in CWE, only 23.6\% had data, with 76.4\% missing; in CAPEC, 56.1\% had data, with 43.9\% missing. The insight is that CAPEC has better data coverage for likelihood information than CWE. 
A final insight from our analysis of the CAPEC database is that the High and Medium likelihood attacks combined to account for 40.5\% of all CAPEC patterns. This suggests a significant portion of documented attack patterns are considered feasible for an attacker to implement and execute.

\textbf{Typical Severity (TS).} This risk feature contains 20.3\% missing or empty values across 5 categories. This risk factor shows an excellent data coverage of 79.7\%, much better than both the CWE likelihood of exploit and the CAPEC likelihood of attack. We also observe that the High and Very High categories combined account for 43.9\% of all attack patterns, indicating a large proportion of severe to extremely severe impact when these attack patterns are implemented. This indicates that nearly half of the documented CAPEC attack patterns are considered highly dangerous. Refer to Figure \ref{fig:ts} for the complete distribution of the categories.





\subsection{Normalization}
\label{sec:normalization}
Stage~3 of our pipeline converts tool outputs into a single, repository-level representation suitable for scoring and cross-repository comparison. Each analyzer emits findings in a different format (query IDs, rule names, locations, and optional traces). Normalization aligns these outputs into a common schema with three goals: (i) reduce duplication across tools, (ii) map findings to standardized weakness identifiers, and (iii) produce a Risk Index for each weakness.

First, we consolidate findings at the repository level: if multiple tools report the same CWE for a repository, we keep a single occurrence of that CWE to prevent duplicate counts.  Second, we map each tool finding to a CWE identifier. CodeQL findings directly emit CWE identifiers as part of their query metadata. Joern findings inherit the CWE label associated with the custom Joern query that triggered the match (Section~\ref{sec:static-code-analysis}). Cisco AI Defender MCP Scanner outputs are mapped to corresponding CWE classes by output category. Third, we compute the CWE Risk Index as defined in Stage~2.

\subsection{Risk Scoring}
\label{sec:risk-scoring}
\subsubsection{Risk Index}
\label{sec:risk-metrics}
In order to calculate a Risk Index, we score each weakness finding \(w\) using a metadata-driven risk model grounded in the standard formulation, \textit{Risk} \(=\) \textit{Likelihood} \(\times\) \textit{Impact}~\cite{stoneburner2002risk}. Each finding \(w\) is associated with a mapping between a CWE identifier \(c_w\) and a CAPEC identifier \(a_w\). We compute the risk only for the cases where a CWE contains related CAPEC identifiers and all required metadata fields; entries with missing or unknown values are removed during preprocessing.

We rely on standardized metadata from the MITRE CAPEC and CWE databases, as defined by their published schemas~\cite{barnum2008common,mitre_cwe_schema}. From CAPEC, we extract \textit{Likelihood\_of\_Attack} and \textit{Typical\_Severity}. From CWE, we extract \textit{Likelihood\_of\_Exploit} and enumerate \textit{Mode\_of\_Introduction} and \textit{Common\_Consequences} fields. We transform these fields from categorical values into corresponding five integer-scaled numerical variables for each \(w\):
\begin{align}
\mathrm{LA}(w) &:= \mathrm{capec.likelihood\_of\_attack}(a_w), \\
\mathrm{LE}(w) &:= \mathrm{cwe.likelihood\_of\_exploit}(c_w), \\
\mathrm{MI}(w) &:= \mathrm{cwe.modes\_of\_introduction\_count}(c_w), \\
\mathrm{CC}(w) &:= \mathrm{cwe.common\_consequences\_count}(c_w), \\
\mathrm{TS}(w) &:= \mathrm{capec.typical\_severity}(a_w).
\end{align}

Using these variables, we define a new \(\mathrm{Likelihood}(w)\), \(\mathrm{Impact}(w)\), and weakness-level risk score \(\mathcal{R}(w)\) as:

\begin{equation}
\label{eq:likelihood_mcp2sos}
\mathrm{Likelihood}(w)=\mathrm{LA}(w)\cdot \mathrm{LE}(w)\cdot \mathrm{MI}(w),
\end{equation}
\begin{equation}
\label{eq:impact_mcp2sos}
\mathrm{Impact}(w)=\mathrm{TS}(w)\cdot \mathrm{CC}(w),
\end{equation}
\begin{equation}
\label{eq:risk_mcp2sos}
\mathcal{R}(w)=\mathrm{Likelihood}(w)\cdot \mathrm{Impact}(w).
\end{equation}

Our design is intentionally conservative: \(\mathrm{Likelihood}(w)\) is the product of CAPEC \(\mathrm{LA}(w)\), CWE \(\mathrm{LE}(w)\), and the mode-of-introduction \(\mathrm{MI}(w)\). This means a finding only receives a high score when it is both plausible to attack and plausible to exploit under the combined CWE–CAPEC metadata. We treat \(\mathrm{MI}(w)\) and \(\mathrm{CC}(w)\) as breadth multipliers: more modes of introduction increase the likelihood term, and more consequence categories increase the impact term.

The Likelihood depends on both the likelihood of a weakness and its associated attack pattern, further amplified by the number of different modes of introduction through which this weakness can be introduced into the codebase for an MCP server. Impact is intensified by the number of common consequences or areas in the software application that will be impacted if an attacker exploits a weakness.
\begin{table*}[t]
\centering
\footnotesize
\setlength{\tabcolsep}{5pt}
\caption{Manual completion for five CWE classes whose CAPEC mappings are missing.}
\label{tab:cwe_manual}
\begin{tabularx}{\textwidth}{|l|p{1.5cm}|p{2.0cm}|X|}
\hline
\textbf{CWE} & \textbf{CAPEC Used} & \textbf{Source} & \textbf{Notes} \\
\hline
CWE-36  & CAPEC-597 & CWE-22 (Path Traversal) &
Official CWE-to-CAPEC linkage exists (Absolute Path Traversal $\to$ CAPEC-597), but CAPEC-597 lacks the required ordinal Likelihood\_Of\_Attack and Severity fields in the current CAPEC release. We impute these from the parent CWE-22 (Path Traversal), which maps to CAPEC-76/126 with comparable attributes (LA=3, TS=5). \\
\hline
CWE-186 & CAPEC-6, CAPEC-15, CAPEC-79 & CWE-185 (Incorrect Regex) &
CWE-186 (Overly Restrictive Regular Expression) has no official CAPEC mapping. We adopt the three CAPEC patterns from its sibling CWE-185 (Incorrect Regular Expression), which shares the same weakness class, adjusting CC from 2 to 1 to reflect a narrower consequence scope. \\
\hline
CWE-639 & --- & CWE-22 (Path Traversal) &
CWE-639 (Authorization Bypass Through User-Controlled Key) has no official CAPEC mapping. We inherit attribute values from CWE-22, as both involve user-controlled input that bypasses access boundaries (LE=3, LA=3, MI=1, CC=3, TS=5). \\
\hline
CWE-863 & CAPEC-114, CAPEC-3 & CWE-285 (Improper Authorization) &
CWE-863 (Incorrect Authorization) is a child of CWE-285. We inherit the two CAPEC mappings from CWE-285 (Authentication Abuse and Using Leading Ghost Characters) with their original attribute values. \\
\hline
CWE-212 & CAPEC-168 & Direct mapping &
Official CWE-to-CAPEC linkage exists (Improper Removal of Sensitive Information $\to$ CAPEC-168), but CAPEC-168 lacks Likelihood\_Of\_Attack, and CWE-212 lacks Likelihood\_Of\_Exploit in the current release. We impute both as 3 (High) based on the attack pattern's description, indicating common exploitability. \\
\hline
\end{tabularx}
\end{table*}
\subsubsection{MCP Server Risk Index}
To prioritize repositories by risk, we aggregate weakness-level risk into per-repository metrics using CWE-level risk and finding frequencies. We denote \(r\) as a repository and \(c\) a single CWE identifier of all \(C\) identifiers. For a repository \(r\), let \(f_r(c)\) denote the frequency of occurrences of \( \forall c \in {C}\) happening in \(r\). Let \(w(c)\) denote the CWE risk used for aggregation, and let \(N_r=\sum_c f_r(c)\) denote the total number of findings in repository \(r\). We define an exposure metric \(R_{\mathrm{exp}}(r)\) as the frequency-weighted sum:
\begin{equation}
\label{eq:repo_exposure}
R_{\mathrm{exp}}(r)=\sum_{c} f_r(c)\cdot w(c).
\end{equation}
We define a severity profile metric \(R_{\mathrm{rms}}(r)\) using a frequency-weighted root-mean-square (RMS) approach, which emphasizes the presence of high-risk CWEs:
\begin{equation}
\label{eq:repo_rrms}
R_{\mathrm{rms}}(r)=\sqrt{\frac{\sum_{c} f_r(c)\cdot w(c)^2}{\sum_{c} f_r(c)}}.
\end{equation}
Finally, we compute a repository-level risk score \(R_{\mathrm{overall}}(r)\) by scaling the severity profile by a log-scaled finding-volume factor. Here, \(N_r=\sum_c f_r(c)\) is the total number of findings in repository \(r\), and we use \(\log_{10}(N_r+1)\) to dampen the influence of repositories with very large numbers of findings:

\begin{equation}
\label{eq:repo_risk}
R_{\mathrm{overall}}(r)=R_{\mathrm{rms}}(r)\cdot \log_{10}(N_r+1).
\end{equation}

Repository risk can vary by total findings $N_r$; a small number of repositories can accumulate exceptionally large \(R_{\mathrm{overall}}(r)\) values (e.g., due to unusually high finding volume \(N_r\) and concentration of high-risk weaknesses \(w(c)\)). Under linear 0 to 100 min-max scaling, these extreme repositories would dominate the range, compressing the remaining repositories into a narrow band near 0, making comparisons and banding uninformative. To reduce the impact of such outliers while preserving the ranking induced by \(R_{\mathrm{overall}}(r)\), we first apply a logarithmic transform, \(\tilde{R}(r)=\ln(R_{\mathrm{overall}}(r))\) and then perform min–max scaling to obtain a normalized score.
\[
R_{\mathrm{norm}}(r)=\frac{\tilde{R}(r)-\tilde{R}_{\min}}{\tilde{R}_{\max}-\tilde{R}_{\min}}\times 100,
\]
where \(\tilde{R}_{\min}=\min_r \tilde{R}(r)\) and \(\tilde{R}_{\max}=\max_r \tilde{R}(r)\). We then assign Very Low/Low/Medium/High/Very High risk bands using fixed 20-point intervals.


\section{Evaluation}
\label{sec:evaluation}
\subsection{Data Sources and Pre-processing}
\label{sec:eval-data}
To obtain our dataset of MCP repositories, we have developed a script using the GitHub Search API. The script queries the \textit{/search/repositories} endpoint with the following conditions \textit{"mcp server stars:>100 pushed:>2025-01-01 language:Python"} and the result is retrieved and sorted by star count. For each repository returned, we record the following metadata: \textit{full\_name}, \textit{stars}, \textit{description}, \textit{url}, \textit{language}, and \textit{updated\_at} to record the repository snapshot time.

We then create the final database by retaining only repositories that actually implement an MCP server and excluding: (i) MCP client-only projects, (ii) benchmark or evaluation harnesses, and (iii) intentionally vulnerable repositories. This yields a dataset of 222 available Python MCP server repositories. The repositories span a wide range of popularity, starting from 101 to 38,167 stars\footnote{Repositories downloaded on January 5, 2026,}. We construct a combined CWE–CAPEC table by linking each CWE to its associated CAPEC attack patterns and keeping only the fields needed for our scoring model. Some CWE–CAPEC entries have missing required fields and we proceed by filling in these gaps by inheriting metadata from the parent category in the taxonomy. A number of 5 CWEs are completed manually out of the 51 as described in Table~\ref{tab:cwe_manual}.

\subsection{Experimental Setup}

We conduct all experiments on an Ubuntu 20.04 server equipped with an AMD EPYC 7313 16-Core processor, 503 GB of RAM, and a NVIDIA A100 GPU with 40 GB of memory. Our static analysis stage uses CodeQL v2.23.8 and Joern v4.0.472. We also use Cisco AI Defender v4.0.3 as a third analyzer on the processed repositories. It includes an LLM-as-a-judge analysis component powered by GPT-4.1. The tool requires setting a parameter, max-files, to indicate how many files should be analysed per repository; we set it to max-files=5. We analyze the main branch for each repository. 

\begin{figure*}[htbp]
\centerline{\includegraphics[width=1.0\textwidth]{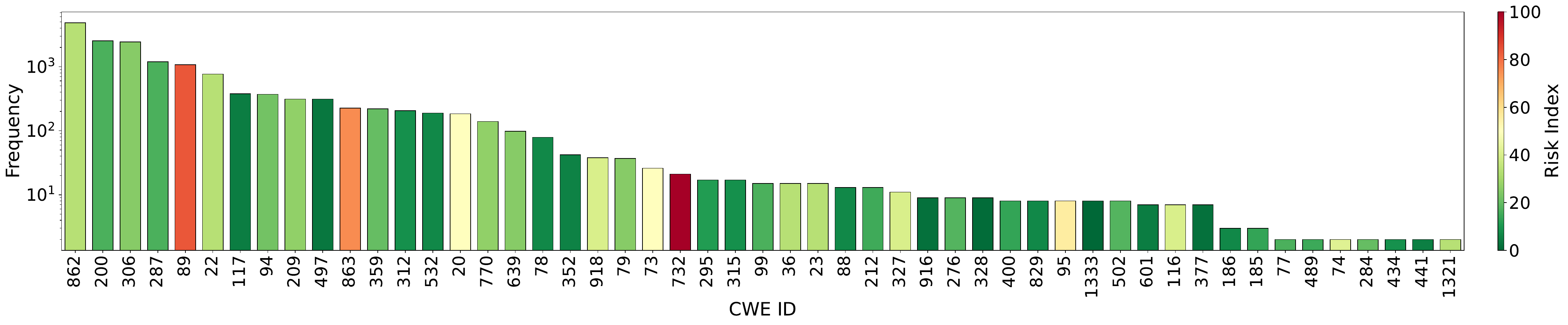}}
\caption{Frequency of each observed CWE class with bar color indicating the Risk Index derived from CWE-CAPEC metadata.}
\label{fig:cwe_frequency_bar}
\end{figure*}

\subsection{Results}
\begin{figure}[htbp]
\centerline{\includegraphics[width=0.3\textwidth]{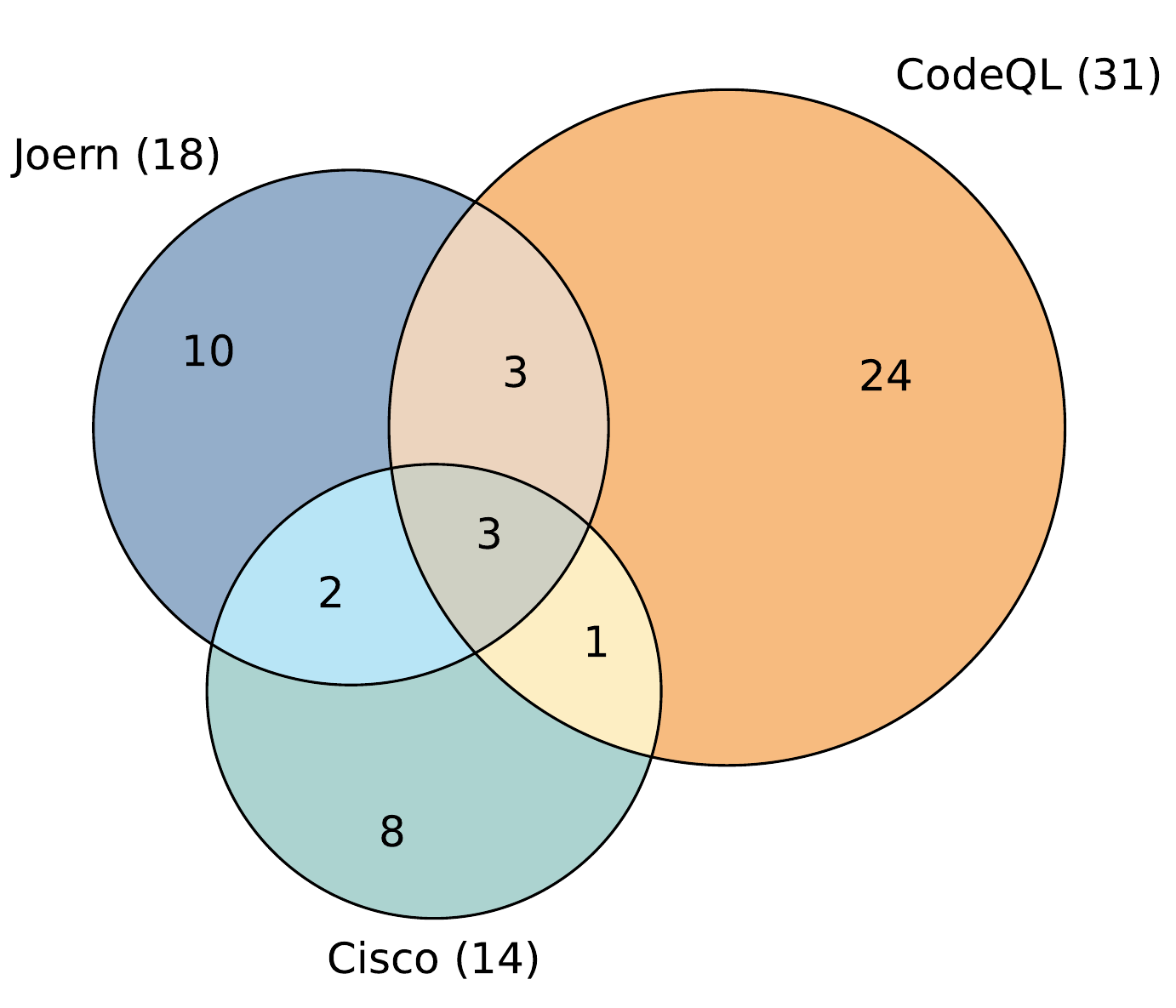}}
\caption{CWE coverage across tools.}
\label{fig:tool_cwe_venn}
\end{figure}

From the 222 repositories obtained in \ref{sec:eval-data}, a total of 191 (86.0\%) exhibit at least one mapped weakness, while the remaining 31 (14.0\%) have no identified weaknesses under our methodology. Figure~\ref{fig:tool_cwe_venn} shows the CWE coverage identified by the code analysis tools, and across all repositories, we identify 51 distinct CWE classes. We next (i) assign a CWE risk score to each mapped finding based on CWE-CAPEC metadata, (ii) aggregate the risk scores to compute a repository-level risk score, (iii) categorize mapped weaknesses into MCP-relevant threat surfaces, and (iv) analyze conditional co-occurrence across these categories to surface common multi-stage exploit chains.

\subsubsection{Risk Assessment}
\label{sec:results:risk_assessment}
\paragraph{CWE Risk Landscape.} For each mapped finding $w$, we compute a Risk Index from CWE-CAPEC metadata and use it as the risk in our repository aggregation (Eq.~\ref{eq:repo_rrms}). Figure~\ref{fig:cwe_frequency_bar} summarizes the resulting CWE landscape. Across the 191 MCP servers with at least one mapped weakness, we identify 15,962 findings spanning 51 distinct CWE classes. The distribution is highly concentrated: the five most common CWEs account for 12,085 findings (75.7\%) and appear in 178/191 repositories (93.2\%). These are CWE-862 (Missing Authorization, 30.4\%), CWE-200 (Exposure of Sensitive Information, 15.9\%), CWE-306 (Missing Authentication, 15.3\%), CWE-287 (Improper Authentication, 7.4\%), and CWE-89 (SQL Injection, 6.7\%). CWE-89 is notable because it is both frequent and high-risk, while the other common classes receive lower to moderate risk scores under our metadata-based model.

The color encoding in Figure~\ref{fig:cwe_frequency_bar} indicates that 41 of 51 CWE classes (80\%) fall in the low Risk Index range, and three in the very-high range. These are, $\mathcal{R}(\text{CWE-732})=100.0$ (Incorrect Permission Assignment for Critical Resource), $\mathcal{R}(\text{CWE-89})=83.3$ (SQL Injection), and $\mathcal{R}(\text{CWE-863})=75.0$ (Incorrect Authorization). In MCP deployments, these three can be especially severe because they match common attacker goals. CWE-732 can allow unintended read/write access to critical server-managed resources (e.g., files, credentials, persisted state) when permissions are too permissive; CWE-89 can allow direct data theft or destructive queries when MCP tool handlers pass unsanitized input into database operations; and CWE-863 can allow privilege escalation when authorization checks exist but are applied incorrectly, letting a client invoke tools or access resources beyond its intended scope.

Conversely, some CWEs with very low Risk Index scores are still important in MCP deployments because the Risk Index is based on high-level, pattern-level metadata rather than how exploitable a specific instance is. For example, CWE-441 (Unintended Proxy or Intermediary, $\mathcal{R}=5.0$) captures the confused-deputy behavior behind prompt injection, where the attacker provides instructions that cause the server to act with its own privileges and unintentionally relay actions or data. In practice, the impact of these failures depends heavily on context, what tools and resources the server can reach, and whether protocol-level access controls are in place, so their true risk may not be fully captured by metadata-only scores.

We emphasize that these Risk Index values are derived from CAPEC’s typical likelihood and severity attributes; MITRE notes that they can vary widely across deployment contexts and are not guaranteed to be accurate for every instance~\cite{mitre-capec-schema-docs-v3.5,barnum2008common}.

\begin{figure}[htbp]
\centerline{\includegraphics[width=0.45\textwidth]{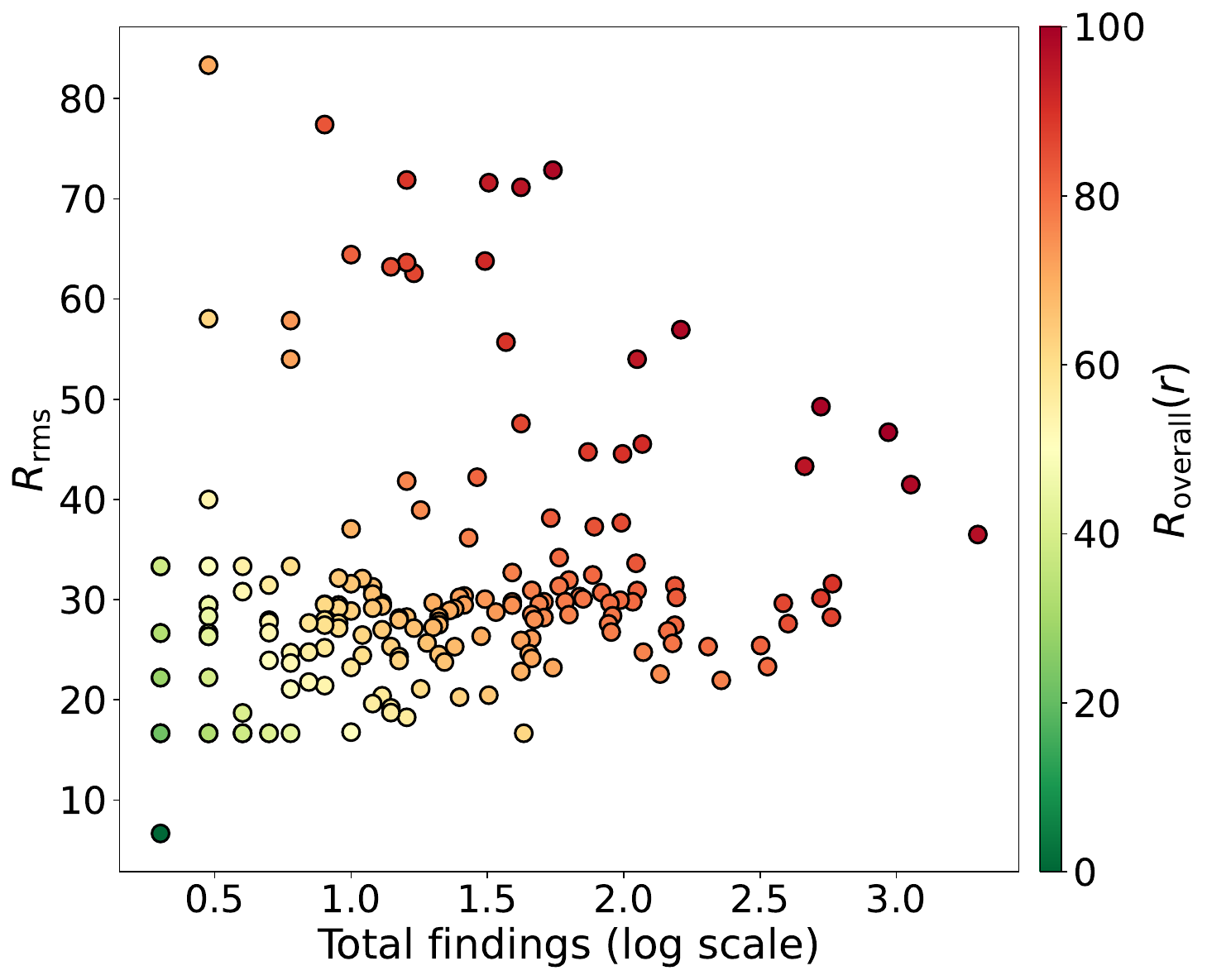}}
\caption{Repository risk versus finding volume.}
\label{fig:risk_per_repo}
\end{figure}

\paragraph{Repository-level Risk.} We next use these CWE-level risks to compute repository-level scores, showing how high-risk weakness classes concentrate within individual MCP servers. Figure~\ref{fig:risk_per_repo} plots each repository by total findings and severity, with color showing the overall repository risk. Total findings measure the breadth of a repository's attack surface, while $R_{\mathrm{rms}}$ measures how much its findings are concentrated in higher-risk CWE classes. 

The scatter reveals two main patterns. First, most repositories cluster around $R_{\mathrm{rms}}\approx 30$ (median 28.9; mean 31.2; 88 repositories in this band) over a wide range of finding counts, meaning many MCP servers share a similar CWE dominated by common, lower-to-moderate-risk classes. These are mainly CWE-862 (Missing Authorization, $\mathcal{R}=33.3$, 154 repos), CWE-306 (Missing Authentication, $\mathcal{R}=25.0$, 129 repos), CWE-200 (Exposure of Sensitive Information, $\mathcal{R}=16.7$, 123 repos), and CWE-22 (Path Traversal, $\mathcal{R}=33.3$, 82 repos), all with severity in the 17 to 33 range.

Second, a smaller subset of repositories has much higher $R_{\mathrm{rms}}$ even with only a moderate number of findings, indicating that their weaknesses are dominated by higher-risk CWE classes. These repositories have a high overall risk without many findings, because those findings sit in the high-risk tail of the CWE distribution. The most dangerous region appears when both factors align: repositories with high $R_{\mathrm{rms}}$ and many findings cluster in the upper-right of Figure~\ref{fig:risk_per_repo} and receive the highest overall risk scores.

\subsubsection{Threat Surfaces and Category}
\label{sec:results:threat_surface}
\begin{figure}[htbp]
\centerline{\includegraphics[width=0.45\textwidth]{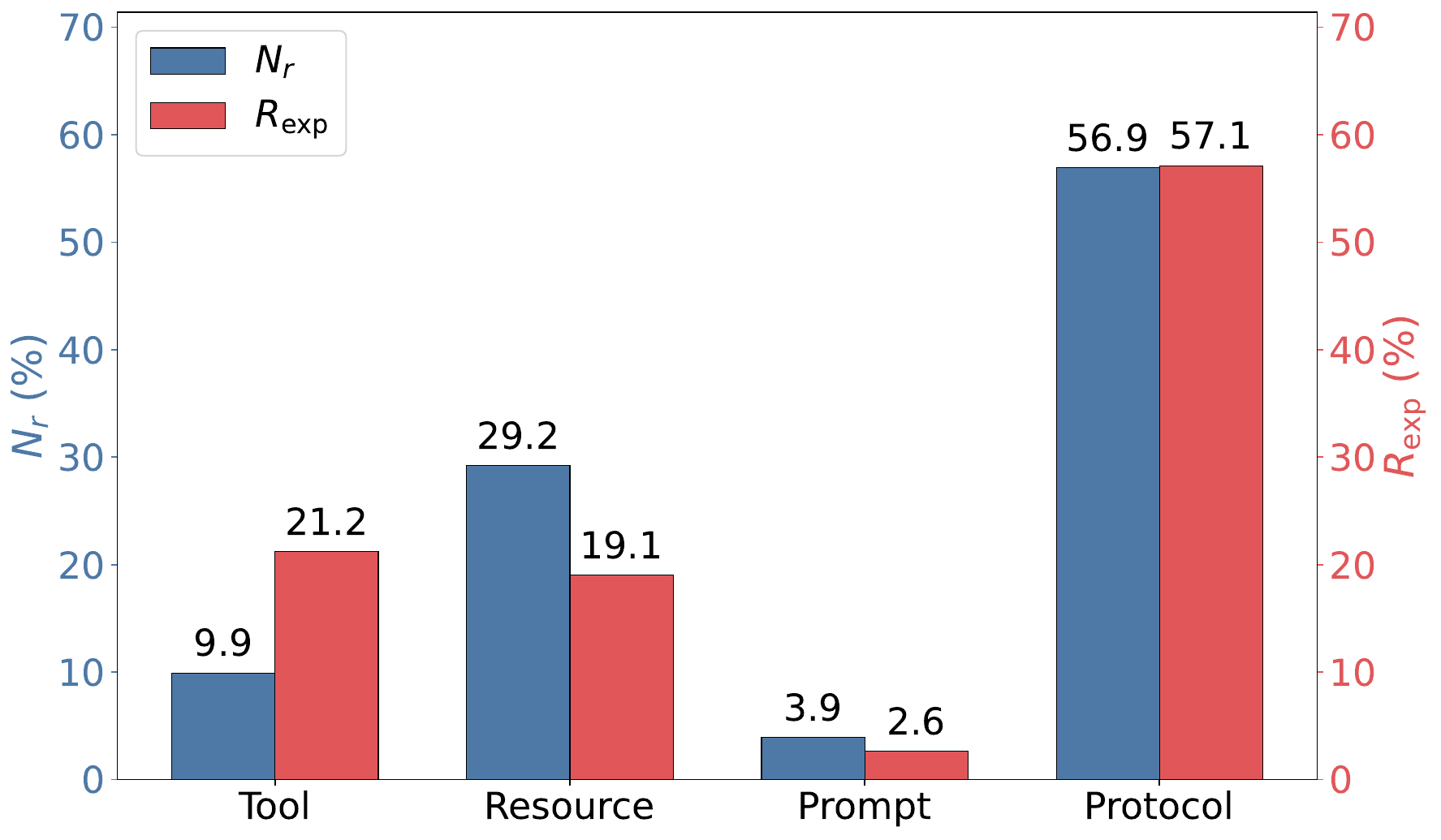}}
\caption{Percentage share of findings ($N_r\%$) and expected exposure ($R_{\mathrm{exp}}\%$) attributed to each MCP threat surface.}
\label{fig:threat_category_bar}
\end{figure}

We next link repository-level risk to the locations where weaknesses arise in MCP servers by mapping each CWE-class finding to one of four threat surfaces: \textbf{Protocol}, \textbf{Resource}, \textbf{Tool}, and \textbf{Prompt}. Figure~\ref{fig:threat_category_bar} shows both (i) the percentage of findings on each surface ($N_r\%$) and (ii) the percentage of expected exposure each surface contributes ($R_{\mathrm{exp}}\%$), so we can directly compare how common a surface is and how much risk it adds. Protocol leads on both measures (56.9\% of findings, 57.1\% of exposure), suggesting that access-control and transport-layer issues act as a reachability multiplier for other weaknesses.  Detailed mappings are present in Appendix Table~\ref{tab:threat-cwe-mapping}.

Resource weaknesses are common (29.2\%) but account for a smaller share of exposure (19.1\%), which is consistent with frequent but varied leakage and misconfiguration problems. Tool weaknesses are less common (9.9\%) but produce a disproportionately large share of exposure (21.2\%), since injection-style tool handlers can have a high impact when they are reachable. Prompt-related weaknesses are rare in our static analysis (3.9\%) and contribute only a small share of exposure (2.6\%), yet they remain strategically important because they can drive Tool and Resource access through confused-deputy behavior when Protocol boundaries are weak.

\begin{figure}[htbp]
\centerline{\includegraphics[width=0.4\textwidth]{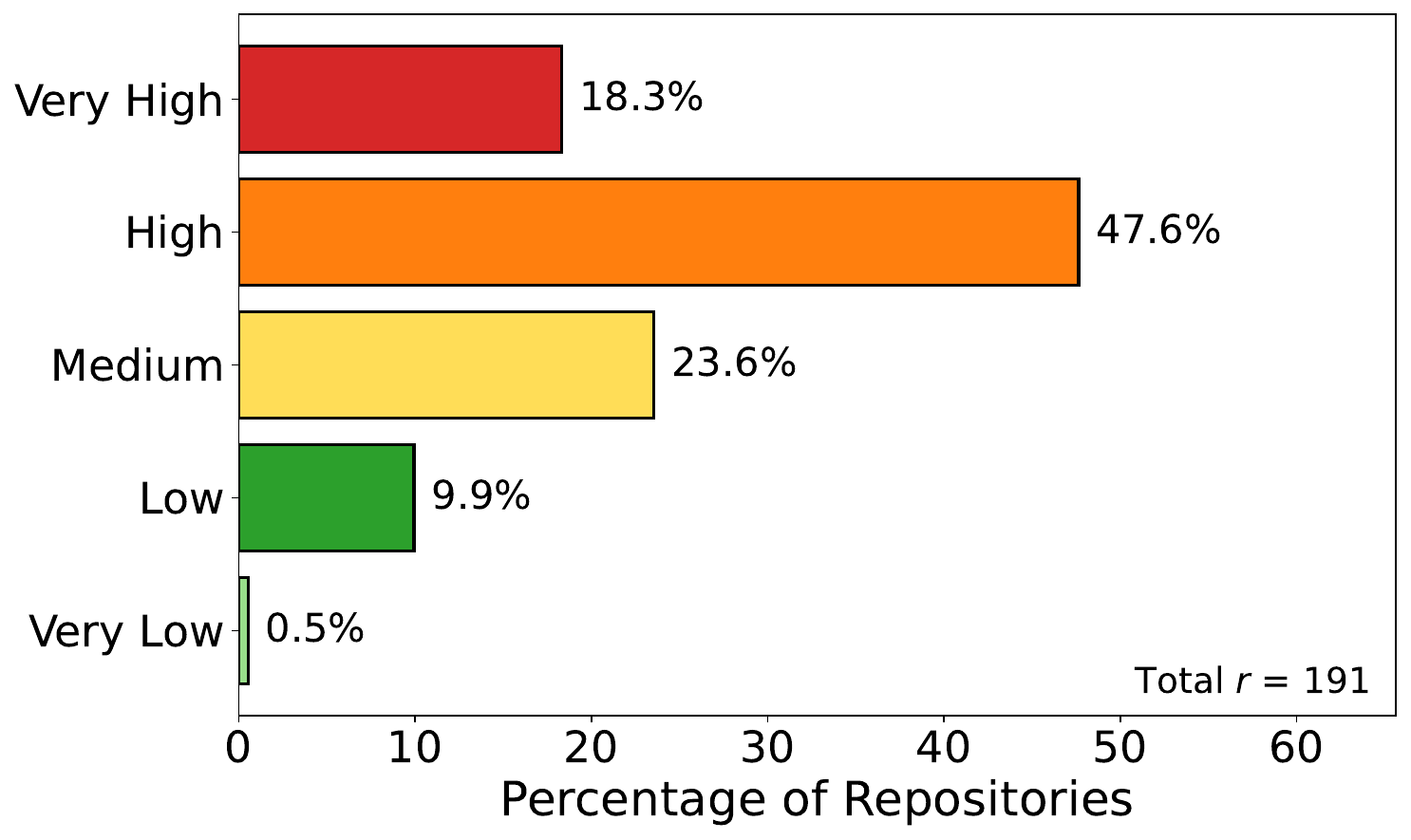}}
\caption{Distribution of repositories across risk levels.}
\label{fig:risk_level_distribution}
\end{figure}

Figure~\ref{fig:risk_level_distribution} shows how these surface-level weaknesses add up to ecosystem-wide risk. Among the 191 scored MCP servers, 47.6\% fall into the High band and 18.3\% into Very High; only 0.5\% are Very Low, with 9.9\% Low and 23.6\% Medium.

\subsubsection{Exploit Chains}
\label{sec:results:cooccur}
\begin{figure}[htbp]
\centerline{\includegraphics[width=0.3\textwidth]{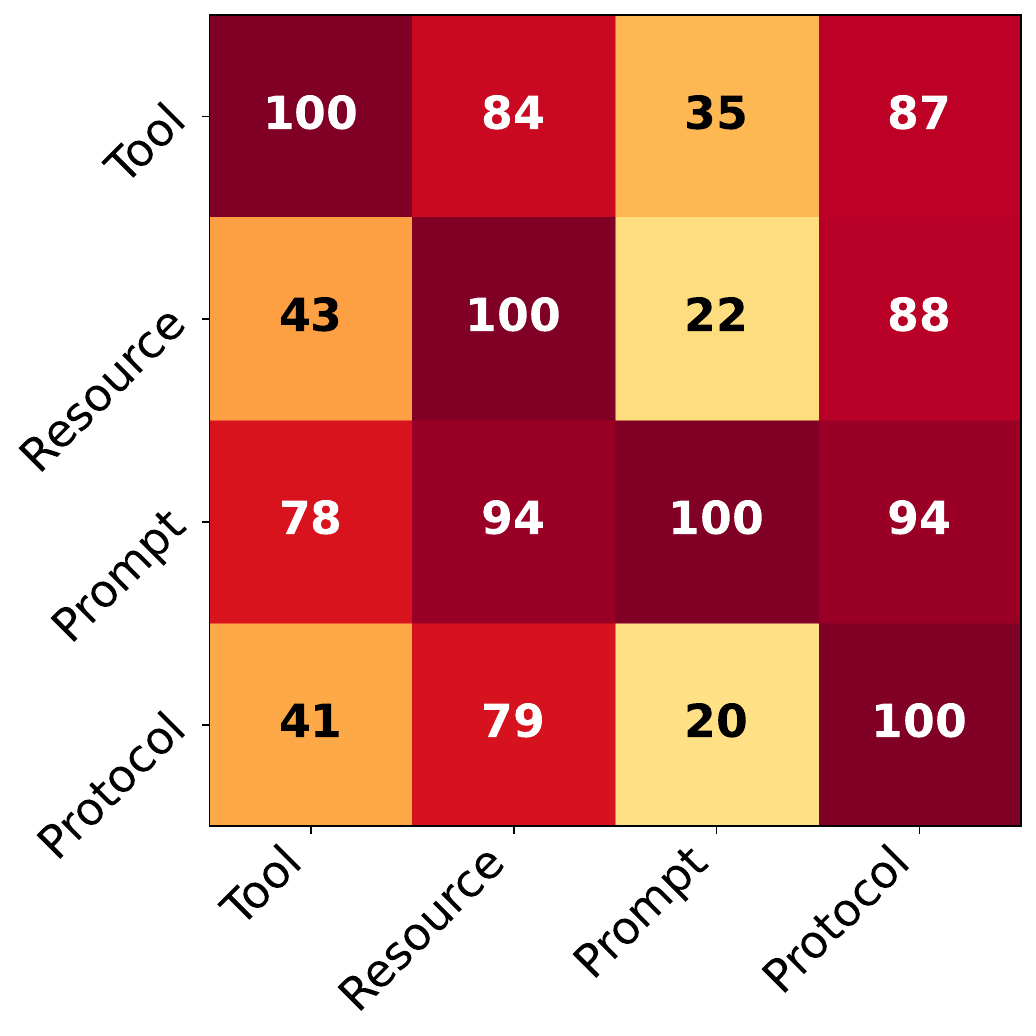}}
\caption{Conditional co-occurrence of threat surfaces.}
\label{fig:threat_cooccurrence_heatmap}
\end{figure}

MCP server weaknesses rarely occur alone; they typically span multiple threat surfaces (Tool, Resource, Prompt, Protocol) and combine into multi-stage exploit chains. To analyze these combinations, Figure~\ref{fig:threat_cooccurrence_heatmap} shows a conditional co-occurrence matrix where each cell $(A,B)$ gives the percentage of repositories that have category $B$ among those that have category $A$. \textbf{Reading convention:} each row fixes category $A$ and columns show how often each category $B$ co-occurs with it.

The strongest asymmetries highlight Protocol weaknesses as a common enabling layer. Repositories with Tool weaknesses also have Protocol weaknesses 87\% of the time, and those with Resource weaknesses have Protocol weaknesses  88\% of the time. In contrast, Protocol weaknesses are much less predictive of Prompt or Tool weaknesses. This pattern is consistent with Protocol issues functioning as reachability. When access control is weak, attackers can exploit downstream Tool and Resource weaknesses, but Protocol issues alone do not necessarily mean there are additional Prompt or Tool weaknesses.

\paragraph{Chain 1: Tool exploitation enabled by Protocol weakness.} Tool weaknesses often appear together with Protocol weaknesses (87\%), and also strongly with Resource weaknesses (84\%). This is a common attack path on MCP servers: weak access control lets attackers reach tool endpoints, and injection-style tool handlers can then be used to steal or change data managed by the server.

\paragraph{Chain 2: Resource exposure via Protocol gaps.}  Resource weaknesses co-occur with Protocol weaknesses in 88\% of cases, suggesting that sensitive data leaks and misconfigurations often hide behind weak access control.
At the same time, (79\%) of repositories with Protocol weaknesses also exhibit Resource issues, indicating that Protocol problems are frequently paired with mismanaged or sensitive resources. This can create direct data-theft chains: protocol weaknesses expand what parts of the system are reachable, and resource weaknesses determine how valuable and sensitive the reachable data is.

\paragraph{Chain 3: Prompt-driven resource theft.} Prompt weaknesses strongly predict both Resource and Protocol weaknesses, at 94\%. This means that when prompt-level confused-deputy failures occur, the system typically already has (i) weak boundaries and (ii) attractive sensitive resources, creating a practical chain where attacker-controlled instructions cause the server to disclose or modify protected data.

\paragraph{Chain 4: Prompt-mediated tool execution.} Prompt weaknesses co-occur with Tool weaknesses in 78\% of cases, meaning prompt manipulation is common in repositories where tool handlers expose injection or execution surfaces. Together with their strong link to Protocol weaknesses, this supports a multi-stage chain where prompt manipulation causes tools to run under weak access control, with tool handlers carrying out the execution or data extraction.

Overall, Figure~\ref{fig:threat_cooccurrence_heatmap} shows that MCP server risk comes less from isolated weaknesses and more from combinations in which Protocol weaknesses frequently appear alongside and help enable Tool/Resource/Prompt exploitation.


\section{Conclusion}
\label{sec:conclusion}
We presented a large-scale weakness assessment of 222 open-source Python MCP servers and found that 191 (86.0\%) exhibit at least one mapped weakness under our methodology. Using CWE-CAPEC-driven risk scoring, 47.6\% of scored repositories fall into the High risk band and 18.3\% into Very High, indicating that elevated risk is common across open-source MCP servers. Our analysis identified 15,962 findings across 51 distinct CWE classes using three analyzers. Mapping these CWEs to four MCP-aligned threat surfaces shows that Protocol weaknesses dominate both prevalence (56.9\% of findings) and risk-weighted exposure (57.1\%). Conditional co-occurrence further shows that threat surfaces rarely appear in isolation; 87\% of repositories with Tool-surface weaknesses also exhibit Protocol weaknesses, and 94\% of repositories with Prompt-surface weaknesses also exhibit Resource-surface weaknesses. We note two limitations. First, our analysis is restricted to Python MCP servers; the weakness distribution may differ for servers written in TypeScript or other languages. Second, our risk scoring relies on CAPEC’s likelihood and severity metadata; these estimates may diverge from realized risk in MCP deployments.

\section{Ethical Considerations}
\label{sec:ethics}
We analyze only publicly available open-source MCP server repositories via offline code inspection and do not interact with deployed systems, users, or data. We report aggregate trends and avoid releasing secrets, exploit-ready payloads, Proof of Concept (PoC) exploits, or any other details that would increase exploitation risk. Following an Ethical Hacking process and Information Security best practices, we reported our findings to the respective companies or open-source groups developing these MCP Servers. During this process, we noticed that not all GitHub repos use the SECURITY.md file, which is the standard GitHub-established way to report vulnerabilities. The information provided to the project owners included a report of vulnerabilities, including file name, code line number, severity, and the PoC exploit used to exploit each vulnerability. We advise the community in general to adopt best security practices and to always implement the SECURITY.md file so that researchers and Bug Bounty hunters have a way to ethically report vulnerabilities.

\section{Open Science}
\label{sec:openscience}
We will release the database metadata (repository list and crawl details), analysis pipeline (preprocessing, CWE-CAPEC mapping, scoring), query artifacts (e.g., Joern/CodeQL configurations), and aggregate tables/scripts sufficient to reproduce all figures. All artifacts will be published in our GitHub repository~\footnote{https://anonymous.4open.science/r/MCP-in-SoS-0947/Readme.MD}.

\bibliographystyle{plain}
\bibliography{references}
\appendix
\section{Appendix}

To connect code-level weaknesses to MCP-specific attack scenarios, we map each of the 51 CWEs to one of four threat categories aligned with the core MCP primitives shown in Table~\ref{tab:threat-cwe-mapping}. \textbf{Tool} threats (11 CWEs) capture weaknesses exploitable through MCP tool invocations, including command injection (CWE-77, 78, 88), SQL injection (CWE-89), code injection (CWE-94, 95), and deserialization (CWE-502). \textbf{Resource} threats (15 CWEs) capture weaknesses in how MCP servers expose and manage data, including path traversal (CWE-22, 23, 36), information exposure (CWE-200, 209, 359), and cleartext storage (CWE-312, 315). \textbf{Prompt} threats (9 CWEs) capture weaknesses exploitable through crafted inputs to the MCP server, including improper input validation (CWE-20), injection (CWE-74, 79), log neutralization (CWE-117), and confused deputy conditions (CWE-441). \textbf{Protocol} threats (16 CWEs) capture weaknesses in the MCP transport and session layer, including missing authentication and authorization (CWE-284, 287, 306, 862, 863), broken cryptography (CWE-327, 328, 916), and denial of service (CWE-400, 770, 1333).

\begin{table*}[htbp]
\centering
\caption{MCP Threat Category to CWE Mapping}
\label{tab:threat-cwe-mapping}
\begin{tabular}{|l|l|}
\hline
\textbf{Threat Category} & \textbf{CWE IDs} \\
\hline
Tool & 77, 78, 88, 89, 94, 95, 99, 434, 502, 829, 1321 \\
\hline
Resource & 22, 23, 36, 73, 200, 209, 212, 276, 312, 315, 359, 377, 497, 532, 732 \\
\hline
Prompt & 20, 74, 79, 116, 117, 185, 186, 441, 601 \\
\hline
Protocol & 284, 287, 295, 306, 327, 328, 352, 400, 489, 639, 770, 862, 863, 916, 918, 1333 \\
\hline
\end{tabular}
\end{table*}

\begin{figure*}[t]
\centering
\centerline{\includegraphics[width=0.7\textwidth]{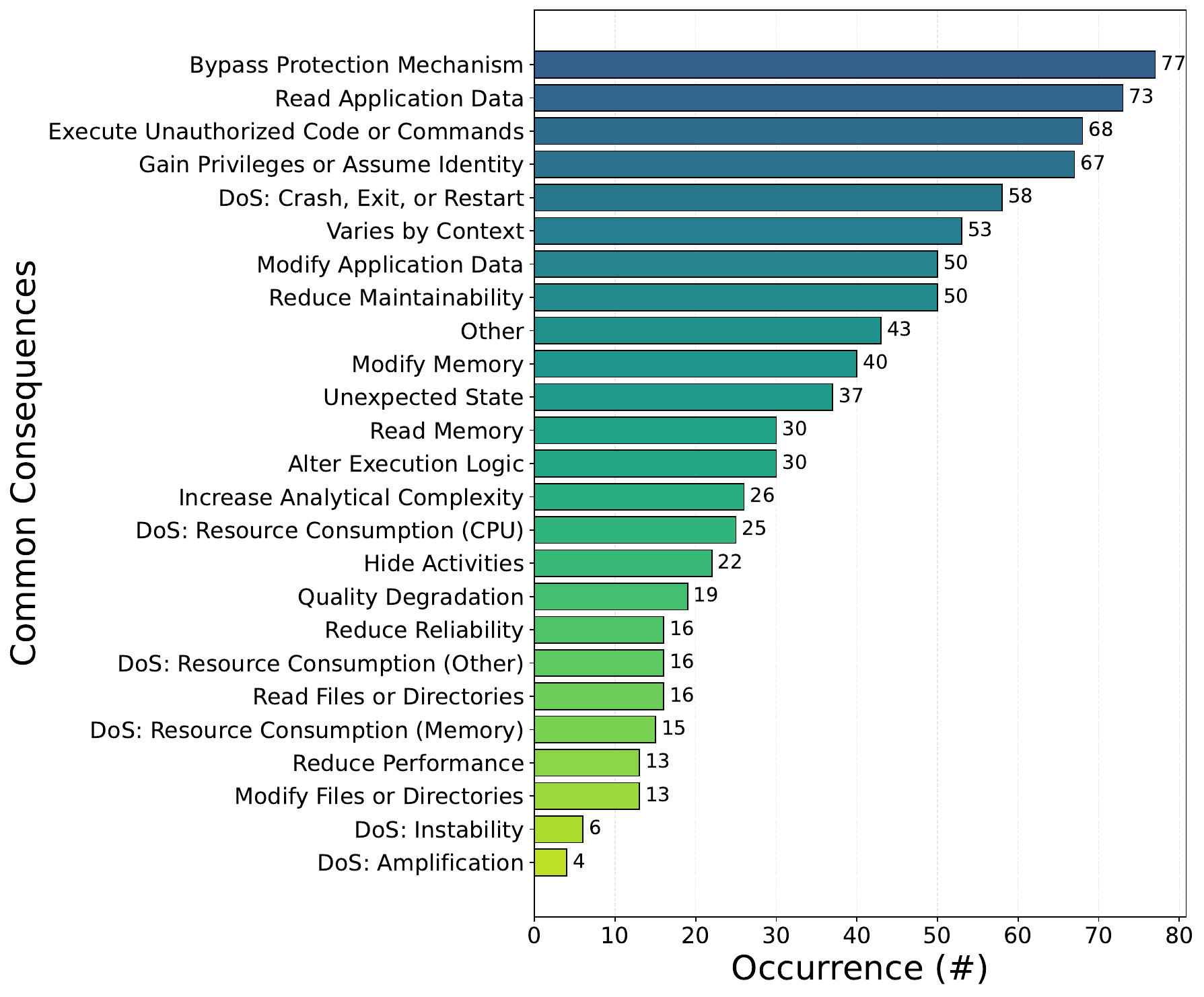}}
\caption{Common Consequences distribution.} 
\label{fig:common_conseq_distribution}
\end{figure*}
\end{document}